\documentclass{XrU2005}
\usepackage{graphicx}

\title{A new XMM-Newton observation of Pictor A radio lobes confirms the non-thermal origin of the X-ray emission}
\author{G. Migliori$^{1,2}$, P. Grandi$^{1}$, G. G. C. Palumbo$^2$, G. Brunetti$^3$, G. Malaguti$^1$, M.  Guainazzi$^4$}
\affil{\small $^1$INAF/IASF Bologna, Italy $^2$Bologna University, Italy $^3$INAF/IRA Bologna, Italy $^4$XMM-Newton SOC Vilspa, Spain  }

\begin{document}


\maketitle

\begin{abstract}
X-ray emission from the eastern radio lobe of the FRII Radio Galaxy Pictor A was 
serendipitously discovered by a short observation of XMM-Newton in 2001. 
The X-ray spectrum, accumulated on a region covering about half of the entire radio lobe, was 
well described by both a thermal model and a power law, making non-univocal the 
physical interpretation. A new XMM-Newton observation performed in 2005 has allowed 
the detection of the X-ray emission from both radio lobes and unambiguously 
revealed its non-thermal origin. The X-ray emission is due to Inverse Compton (IC) 
of the cosmic microwave background photons by relativistic electrons in the lobe. 
We confirm the discrepancy between the magnetic field, as deduced from the 
comparison of the IC X-ray and radio fluxes, and the equipartition value.
\end{abstract}

\section{Introduction}

Pictor A is a nearby ($z$=0.035) double lobe radio source with a Fanaroff Riley II 
(FRII) morphology. VLA observations (Perley et al. 1997) show two nearly circular 
radio lobes with hot spots and a faint radio jet connecting the nucleus to the 
west hot spot. The spectral indices are fairly uniform throughout each 
lobe $<\alpha_{\rm R}>=0.8$. XMM-Newton discovered X-ray radiation 
associated to the extended radio lobes (Grandi et al. 2003). In spite of 
the relatively short exposure time ($\sim15$ ks), and the very high background, a 
quite good spectrum was obtained from a circular region centered on the east lobe. 
However, the data interpretation was not univocal, as the X-ray spectrum could 
be described by both a power law ($\alpha_{\rm X}=0.6\pm0.2$) and a thermal model 
($kT\sim5$ keV). The presence of an extended emission, most likely to be 
ascribed to non-thermal processes, has been recently confirmed (Hardcastle \& Croston 2005). 
To definitely solve the ambiguity on the 
nature of the radio-lobes X-ray emission, a new XMM-Newton 
50 ks observation was performed on 14 January 2005. Here we present the preliminary results.

\begin{figure}[h]
\vspace{7cm}
\caption{XMM/MOS1 image (0.2-10 keV) of Pictor A observed on 14 January 2005.
Several components are visible: the bright nucleus, the west hot spot, the jet, and the two lobes.
The green and the pink circles (excluding the nuclear and the jet contributions), represent the extraction regions 
of the east and west lobe spectra, respectively.}
\end{figure}

\section{Spectral analysis }

\begin{table}
\centering
\caption{East lobe spectral fit results}\vspace{0.1cm}
\begin{tabular}{|lr|lr|} \hline
              &                       &             &                       \\
\multicolumn{2}{|c|}{Power Law} &
\multicolumn{2}{|c|}{Thermal Emission} \\ 
              &                       &             &                       \\ \hline
              &                       &             &                       \\
 $\Gamma$     &$1.80^{+0.17}_{-0.14}$ & $kT$(keV)   &$4.94^{+3.01}_{-1.47}$  \\
              &                       &             &                       \\
$F^{(a)}$     &$1.09^{+0.21}_{-0.12}$ & $F^{(a)}$   &$1.00^{+0.01}_{-0.01}$  \\
              &                       &             &                       \\
$\chi^2/$dof  & 19/23                 &$\chi^2/$dof &28/23 \\
              &                       &             &                       \\ \hline
\multicolumn{4}{l}{$(a)$: 2--10 keV flux in $10^{-13}$ erg cm$^{-2}$ s$^{-1}$}\\
\end{tabular}
\end{table}

\begin{table}
\centering
\caption{West lobe spectral fit results}\vspace{0.1cm}
\begin{tabular}{|l r|l r|}\hline
              &                       &             &                       \\
\multicolumn{2}{|c|}{Power Law} &
\multicolumn{2}{|c|}{Thermal Emission} \\
              &                       &             &                       \\ \hline
              &                       &             &                       \\
$\Gamma$      &$1.75^{+ 0.17}_{-0.17}$ &kT$(keV)$     & $5.59^{+3.18}_{-1.62}$  \\
              &                       &             &                       \\
$F^{(a)}$     &$1.58^{+ 0.16}_{-0.16}$ &$F^{(a)}$     & $1.48^{+0.12}_{-0.12}$   \\
              &                       &             &                       \\
$\chi^2/$dof  & 34/30                  & $\chi^2/$dof & 42/30 \\ 
              &                       &             &                       \\ \hline
\multicolumn{4}{l}{$(a)$: 2--10 keV flux in $10^{-13}$ erg cm$^{-2}$ s$^{-1}$}\\
\end{tabular}
\end{table}

Figure 1 shows the observed MOS1 image (0.2-10 keV) of Pictor A. 
The X-ray counterparts of radio lobes are clearly visible as asymmetric diffuse 
emission around the bright nucleus. Unlike the previous XMM-Newton observation, 
performed in 2001, spectra of good quality can be extracted from both lobes. This 
has allowed a spatially resolved X--ray spectral analysis.
The green circle indicated in Figure 1 represents the selected accumulation region 
for the east lobe spectrum. The spectrum of the west lobe was instead extracted from the pink circle, after subtraction 
of the nuclear and the jet contributions (black circle and rectangle, respectively).

Following the previous work (Grandi et al. 2003) we tested both a power law and a thermal model to fit the data. For both lobes, a 
non-thermal emission is preferred to a thermal process, as shown in Tables 1 and 2. This appears also evident in Figure 2 where 
the two spectral models to the east lobe data are compared. The residuals result clearly larger for the thermal model fit. The 
comparison between Table 1 and 2 indicates that the X-ray photon index value is $\Gamma\sim1.8$ ($\alpha_{\rm X}=0.8$) 
independently of the position of the extraction region. The X-ray spectral index, $\alpha_{\rm X}$ is thus in very good agreement 
with the radio slope ($\alpha_{\rm R}\sim0.8$). Moreover, the non-thermal X-ray flux of the east lobe is in agreement with the 
previous value reported in Grandi et al. (2003).

\begin{figure}
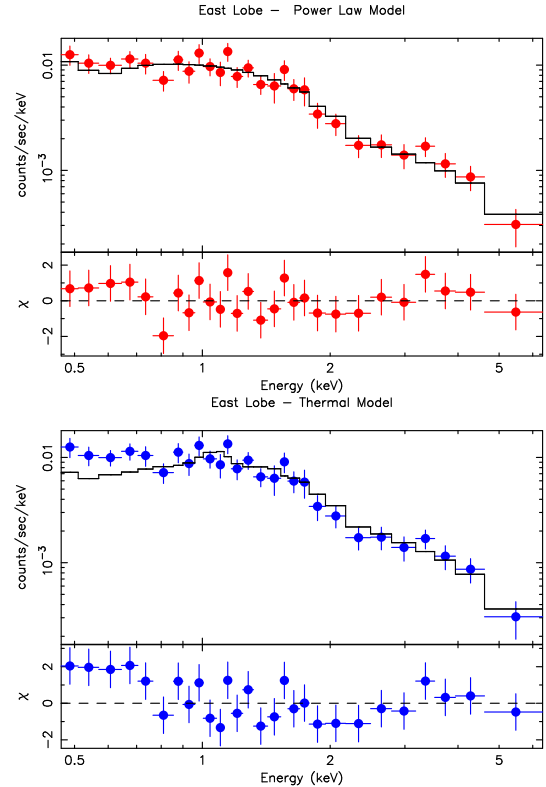

\centering
\includegraphics[scale=0.3,angle=-90]{lobestpl.cps}\\
\includegraphics[scale=0.3,angle=-90]{lobesthermal.cps}
\caption{XMM/MOS1 spectrum of the east lobe: a power law (upper panel) is clearly preferred to a thermal model (bottom panel).}
\end{figure}

\section{Discussion}

This new observation solves the ambiguity on the nature of the extended X-ray emission. Radio emitting electrons in the 
lobes upscatter local cosmic microwave background, producing the observed X-ray emission. 
We assume $k=1$, where $k$ is the ratio between electron and proton energy density in the lobes, 
and an electron spectrum $N(\gamma) \propto \gamma^{-(2\alpha_{\rm R}+1)}$, with a low energy cut-off $\gamma_{\rm min}=50$.
Following Brunetti et al. (1997) and Blumenthal \& Gould (1970), we find $B_{\rm EQ}/B_{\rm IC} \simeq 2.7$, which implies 
a ratio between the particle and magnetic field energy density equal to (Brunetti 2004): 
$(\omega_{\rm (e+p)}/\omega_{\rm B})\sim[2/(\alpha_{\rm R}+1)](B_{\rm EQ}/B_{\rm IC})^{(3+\alpha_{\rm R})} \sim 50$,
thus showing a strong particle dominance in the radio lobes.


\begin{thebibliography}{}
\bibitem[Blu(1970)]{Bl:2004} Blumnethal., G. R. \& Gould, R.J. 1970, Rev. Mod. Phys., 42, 237
\bibitem[Brunetti et al (1997)] {BR:1997} Brunetti G., Setti, G. \& Comastri, A., 1997, A\&A, 325, 898
\bibitem[Brunetti(2004)]{BR:2004} Brunetti G. 2004, in The Role of VLBI in Astrophysics, Astrometry and Geodesy, Mantovani \& kus eds, p. 29.
\bibitem[Grandi(2003)]{PG:2003} Grandi, P. et al. 2003, \apj, 586, 123
\bibitem[Hardcastle (2005)]{HC:2005} Hardcastle. M., J. \& Croston, J. H. 2005, \mnras, 363, 649
\bibitem[Miley(1980)]{MI:1980} Miley, G., 1980 \araa, 18, 165
\bibitem[Perley(1997)]{PE:1997} Perley R.A., Roser H.J. $\&$ Meisenheimer K. 1997 \aap, 328, 12
\end{thebibliography}
\end{document}